\documentclass[11pt]{article}
\usepackage{graphicx,floatflt,amssymb,epsf,rotate}
\usepackage{epsfig} 
\newcommand{\be}{\begin{eqnarray}}
\newcommand{\ee}{\end{eqnarray}}

\def\be{\begin{equation}}
\def\ee{\end{equation}}
\def\gs{\mathrel{
   \rlap{\raise 0.511ex \hbox{$>$}}{\lower 0.511ex \hbox{$\sim$}}}}
\def\ls{\mathrel{
   \rlap{\raise 0.511ex \hbox{$<$}}{\lower 0.511ex \hbox{$\sim$}}}}

\newcommand{\ba}{\begin{array}{c}}
\newcommand{\baz}{\begin{array}{cc}}
\newcommand{\barrr}{\begin{array}{rrr}}
\newcommand{\bad}{\begin{array}{ccc}}
\newcommand{\bav}{\begin{array}{cccc}}
\newcommand{\baf}{\begin{array}{ccccc}}
\newcommand{\bea}{\begin{equation} \begin{array}{c}}
\newcommand{\eea}{ \end{array} \end{equation}}
\newcommand{\ea}{\end{array}}

\begin{document} 
\begin{flushright} 
RECAPP-HRI-2010-011\\
\end{flushright} 
 
\vskip 30pt 
 
\begin{center} 
{\Large \bf An SO(10) model with adjoint fermions for double seesaw neutrino 
masses}\\
\vspace*{1cm} 
\renewcommand{\thefootnote}{\arabic{footnote}} 
{ {\sf Joydeep Chakrabortty 
\footnote{E-mail address: joydeep@hri.res.in}$^\dagger$},
{Srubabati Goswami
$\footnote{E-mail address:
sruba@prl.res.in}^{\ddagger}$}},
{Amitava Raychaudhuri
\footnote{E-mail address:
raychaud@hri.res.in}$^{\dagger\star}$}  \\

\vskip 10pt
\small$\phantom{i}^{\dagger}${\em Harish-Chandra Research Institute,\\
Chhatnag Road, Jhunsi, Allahabad - 211019, India }   \\

\vspace{10pt}

\small$\phantom{i}^{\ddagger}${\em Physical Research Laboratory,\\
 Navarangpura, Ahmedabad - 380009, India}\\

\vskip 10 pt

\small$\phantom{i}^{\star}${\em Department of Physics, University
of Calcutta,\\ 92 Acharya Prafulla Chandra Road, Kolkata -
700009, India }
\\

\normalsize 
\end{center}

\begin{abstract}
An $SO(10)$ model where the $10_H$ and $120_H$ representations
are used for generating fermion masses is quite predictive,
though due to the absence of $SU(2)_{L,R}$ triplet/singlet fields it
cannot give rise to neutrino masses through the usual type-I or
type-II seesaw mechanisms.  In this paper for neutrino masses we
propose an extension of such an $SO(10)$ model by adding fermions
in the adjoint representation (${45}_F$) and a symmetry
breaking scalar $\overline{16}_H$. The $\overline{16}_H$ couples
the adjoint fermions to the standard fermions in
${16}_F$ and induces neutrino masses through the `double
seesaw' mechanism. In order to enhance the predictivity of the
model we impose $\mu-\tau$ flavour symmetry on the Yukawa
matrices for $10_H$ and $\overline{16}_H$ whereas for the $120_H$ it is
assumed to be antisymmetric. We discuss the conditions that the
mass matrices must obey so that the model can reproduce the
tri-bimaximal mixing pattern.

\end{abstract}
\pagebreak

\renewcommand{\thesection}{\Roman{section}} 
\setcounter{footnote}{0} 
\renewcommand{\thefootnote}{\arabic{footnote}} 
\noindent

\section{Introduction}

A number of experiments with solar, atmospheric, reactor and
accelerator neutrinos have now unambiguously established that
these elusive particles are massive.  In addition, the data imply
one small and two large mixing angles in complete contrast with
the quark sector where all three mixing angles are small.  In
Table \ref{t:fits} we present the best-fit values and 3$\sigma$
ranges of neutrino oscillation parameters as obtained from the
global oscillation analysis \cite{thomas}.  These values are
close to the so called tri-bimaximal mixing pattern \cite{tribi}
which implies $\sin^2\theta_{12} = 1/3$, $\sin^2\theta_{23} =
1/2$ and $\sin^2\theta_{13} =0$.

Since in the Standard Model (SM)  neutrinos are massless this
compels one to transcend  beyond the realms of the SM.  There are
also  several theoretical motivations for going beyond the SM,
one of which is that the SM is a product of three gauge groups 
and so involves three independent couplings. A Grand Unified
Theory (GUT), which is a theory of strong and electroweak
interactions based on a  single gauge group \cite{Georgi1}, aims
to unify the three forces with a single coupling constant
\cite{Georgi2}.  It also unifies the matter fields by placing the
quarks and leptons in the same irreducible representation of the
underlying gauge group \cite{Pati}.  Since GUTs aim to unify
quarks and leptons it is a challenge to reconcile the large
mixings in the lepton sector with the small mixings in the quark
sector. The issue of fermion masses and mixing in the context of
GUTs has received much attention from this perspective.

Several GUT models based on gauge symmetries such as $SU(5)$,
$SO(10)$, and $E(6)$ have been proposed and studied extensively.
The minimal GUT group which has the same rank as 
${\mathcal G}_{SM} \equiv SU(3) \otimes SU(2)_L \otimes U(1)_Y$ is
$SU(5)$ \cite{Georgi1}.  $SU(5)$ requires two different
representations ($\bar{5} + 10$) to accommodate all the fermions
of one generation. Moreover the minimal model does not achieve
gauge coupling unification neither does it allow a neutrino mass.
On the other hand, $SO(10)$ GUT has the  feature  of unifying all
quarks and leptons within its 16-dimensional spinor
representation \cite{so10}. This accounts for the 15 SM fermions
and a right-handed neutrino and allows a natural implementation
of the seesaw mechanism \cite{Seesaw}. It has been shown in
a number of papers that renormalizable $SO(10)$ -- with and without
supersymmetry (SUSY) -- is quite  predictive and powerful in
constraining fermion mass patterns because of the underlying
$SU(4)_c$ symmetry which relates the quark and lepton Yukawa
couplings.  In $SO(10)$, $16\otimes 16=10 \oplus 120 \oplus 126$
and so Higgs fields giving mass to the $16_F$ can reside in the
$10_H$, $120_H$ and $\overline{126}_H$ representations.
Obtaining correct masses for the quarks and the charged leptons
requires at least two Higgs multiplets.  
It has been noted, for example in \cite{goran}, that any one of the
combinations $(10_H, 120_H)$, $(10_H, \overline{126}_H)$, or
$(120_H, \overline{126}_H)$ can, in principle, be utilized.
Among
these the model with $10_H$ and $\overline{126}_H$ has been
extensively considered as the  most successful candidate for the
minimal $SO(10)$ GUT \cite{babu-mohapatra}. $\overline{126}_H$
contains colour singlet submultiplets which transform as a
triplet under $SU(2)_{L}$ and a singlet under $SU(2)_{R}$ or {\em
vice versa}; these are the cornerstones of the seesaw mechanism
\cite{Seesaw}. Both type-I (mediated through singlets
\cite{Seesaw}) and type-II (mediated through scalar triplets
\cite{seesaw2}) seesaw have been examined for both supersymmetric
\cite{10+126-susy} and non-supersymmetric \cite{10+126-nosusy}
cases.  The $\overline{126}_H$ relates the Majorana mass of the
neutrinos to the Dirac mass as well as other charged fermion
masses making the model predictive.  It is also possible and in
some cases advantageous to include all the three Higgs
representations \cite{allthree,anjan-ketan}.  The model with
$10_H + 120_H$ \cite{10+120, Lavoura:2006dv}, on the other hand,
does not have the requisite scalars to lead to neutrino masses
through the seesaw mechanism. Here, neutrino mass can be obtained
at two loop through the radiative seesaw mechanism due to Witten
\cite{witten} by adding  $16_H+\overline{16}_H$ multiplets.  This
model has been studied in \cite{goran-bajc-prl} and it was shown
that under plausible assumptions it predicts $b-\tau$
unification, natural occurrence of large leptonic and small quark
mixing and large value for the atmospheric mixing angle.
However, the radiative seesaw runs into difficulty with
low-energy SUSY although it works well in the context of split
SUSY \cite{goran-split}.  Moreover, as has been shown in
\cite{Lavoura:2006dv} the SUSY $SO(10)$ model containing $10_H$
and $120_H$ cannot reproduce the charged fermion masses
correctly. On the other hand in non-SUSY $SO(10)$ the two-loop
neutrino mass is very small.

\begin{table}
\begin{center}
\begin{tabular}{c|c|c}\hline\hline
& best fit & $3\sigma$ range   \\\hline
$\Delta m_{21}^2$ [$10^{-5}~{\rm eV}^2$]  & 7.59 & 7.03 - 8.27 \\
$|\Delta m_{31}^2|$ [$10^{-3}~{\rm eV}^2$]  & 2.40 & 2.07 - 2.75 \\\hline
$\sin^2\theta_{12}$ & 0.318 & 0.27 - 0.38 \\
$\sin^2\theta_{23}$ & 0.50 & 0.36 - 0.67 \\
$\sin^2\theta_{13}$ & 0.013 & $\leq$ 0.053 \\\hline\hline
\end{tabular}
\end{center}
\caption{The  best-fit values and the 3$\sigma$ ranges
of neutrino mass and mixing parameters as obtained from a global analysis of 
oscillation data \cite{thomas}.
$\Delta m_{ij}^2 = m_i^2 - m_j^2$.
}
\label{t:fits}
\end{table}

In this paper we consider the generation of neutrino masses in
the $10_H + 120_H$ model embellished with a $\overline{16}_H$ by adding
fermions belonging to the adjoint representation ($45_F$) of $SO(10)$.
Such fermions couple to the usual sixteen-plet of quarks and
leptons {\em via} the $\overline{16}_H$ and can give rise to neutrino masses
through the `double seesaw' mechanism.  In models with $10_H +
120_H$ this can serve as an alternative option for generating
small neutrino masses\footnote{It is also possible to get a
double seesaw type mass matrix using singlet fields
\cite{double-seesaw}.}.  Fermions in the triplet adjoint representation
of $SU(2)_L$ are also considered in the so called type-III \cite{TypeIII}
seesaw mechanism. Such models have become quite popular in the
context of $SU(5)$ GUTs \cite{perez-su5}. 
$SU(2)_L$ triplet fermions 
fit naturally into the 24-dimensional representation of $SU(5)$
and can cure two main problems of these theories, {\em viz.}
generation of neutrino masses and unification of gauge couplings.
The latter requires the mass of the fermionic triplets to be
$\sim$ ${\cal O}$(1 TeV) making the model testable at the LHC
\cite{goran-triplet}. 
Presence of adjoint fermions in the context of left-right
symmetric models has been considered in \cite{perezlr} and
generation of neutrino masses and possible collider signatures
were discussed.  From this point of view our model can also be
considered as a generalization of type-III seesaw for $SO(10)$.
However as in LR symmetric models the mechanism of mass
generation here is actually the `double seesaw' mechanism.

We discuss the conditions which the Yukawa coupling matrices should
satisfy for the model to have predictive power.  This requires
ascribing some additional flavour symmetry to the model which
we choose to be the  generalized $\mu-\tau$ symmetry that has
been considered widely for explaining the neutrino mixing angles
\cite{mu-tau}.  It predicts  $\theta_{23}$ to be $\pi/4$
which is the best-fit value of this angle from global fits. In
addition it implies $\theta_{13}=0$ which is also consistent
with the data.  Small deviation from  these exact values may be
generated by breaking the $\mu-\tau$ symmetry  by a small amount.
Combining $\mu-\tau$ flavour symmetry with GUTs has been
considered in the case of $SU(5)$ in \cite{mohapatra-nasri} and
also for $SO(10)$ \cite{anjan-ketan}.  Here we impose $\mu-\tau$
symmetry on the Yukawa matrix for the $10_H$ and
$\overline{16}_H$ whereas the one for $120_H$ is taken to be
antisymmetric.  We also impose a parity symmetry  leading to
Hermitian Yukawa matrices.  Thus we consider the model $SO(10)
\otimes Z_2^{(\mu-\tau)} \otimes Z_2^{\cal P}$
\cite{anjan-ketan}.  Imposition of these two symmetries help in
reducing the number of unknown parameters in the Yukawa sector.
In addition, we make an  ansatz relating the effective $\nu_R$
mass matrix arising due to the inclusion of adjoint fermions with
the Yukawa matrix for $10_H$.  As a result the light neutrino
mass matrix after seesaw mechanism obtains a simple form and can
be written as a sum of two contributions.  It turns out that with
the above choice the neutrino mass matrix is $\mu-\tau$ symmetric
so that one immediately gets $\theta_{13} =0$ and $\theta_{23} =
\pi/4$. It is straight-forward to get the prediction for the
neutrino masses and $\theta_{12}$ and obtain the conditions on
the parameters such that tri-bimaximal mixing is obtained.  We
also present the limiting values when one of the two
contributions dominates.  With the above set of assumptions one
can get masses and mixing angles consistent with those presented
in Table \ref{t:fits}.

The plan of the paper is as follows. In the next section we
discuss the model. In section III we compute the evolution of the
gauge couplings in the context of this model and obtain the
range of the intermediate as well as unification scales.  In
section IV we discuss the neutrino mass matrix. Finally in section V
we impose $\mu-\tau$ symmetry and obtain predictions for neutrino
masses and mixing angles. We end with the conclusions.

\section{The Model}
We explore an $SO(10)$ model where the three fermion families acquire
mass through the $10_H$ and/or ${120}_H$.  The model  also
includes additional fermion multiplets in the $SO(10)$ adjoint
representation, 45$_F$, and a $\overline{16}_H$.

In this model the Yukawa terms  for the fermions can be expressed
as:
\begin{equation}
\mathcal{L}=Y_{10}16_F 16_F 10_H +Y_{120}16_F 16_F {120}_H .
\end{equation}
In general, $Y_{10}$ is a complex symmetric matrix while
$Y_{120}$ is complex antisymmetric. When the $10_H$ and $120_H$
scalars obtain their vacuum expectation values ({\em vevs})
quarks and leptons obtain masses which can be represented as:
\begin{eqnarray}
\label{matrices}
m_d &=& M_0 + i M_2, \;\; 
m_u = c_0 M_0 + i c_2 M_2, 
\nonumber\\
m_l&=& M_0 + i c_3 M_2, \;\; 
m_D = c_0 M_0 + i c_4 M_2.
\end{eqnarray}
Above, $m_d$ ($m_u$) denotes the mass matrix for the $d$-type
($u$-type) quarks, $m_l$ is the charged lepton mass matrix,
whereas $m_D$ is the Dirac mass matrix of the neutrinos.  The
matrices $M_0$ and $M_2$ are proportional to $Y_{10}$ and
$Y_{120}$ respectively.
\begin{equation}
M_{0}=M_{0}^T, \;\; M_{2}= - M_{2}^T.
\label{m10120} 
\end{equation}
$c_0, c_2, c_3$, and $c_4$ are constants fixed by Clebsch-Gordan
(CG) coefficients and {\em vev} ratios which are taken to be
real.  We impose a generalized parity symmetry and make
appropriate choices of the {\em vev}s \cite{parity} which make
$M_0$ and $M_2$ real thereby reducing the number of free
parameters and ensuring the hermiticity of the mass matrices in
eq. (\ref{matrices}).

For neutrinos the above implies the presence of only the Dirac
mass term which cannot reproduce the correct neutrino mass pattern
\cite{goran-bajc-prl}.  Since the $\overline{126}_H$ field is not
present the type-I and type-II seesaw mass terms are absent in
this model.  One can  of course generate the neutrino mass
through the Witten mechanism of radiative seesaw \cite{witten}
but then for non-SUSY $SO(10)$ such contributions are too small
\cite{goran-bajc-prl}. 

In this work we propose a new mechanism to generate a neutrino mass in 
a non-SUSY $SO(10)$ with $10_H$ and $120_H$. 
We introduce additional matter multiplets 
($45_F$) which belong to the  adjoint representation of $SO(10)$.
Note that this is similar to the so called type-III seesaw
mechanism where one adds additional matter fields in the adjoint
representation. However, as we will see, the neutrino mass is
generated here through the `double seesaw' mechanism.  
$SO(10)$ breaks to the SM through two intermediate steps:
\begin{equation}
SO(10)\stackrel{M_X}{\longrightarrow} SU(4)_c \otimes SU(2)_L
\otimes SU(2)_R \stackrel{M_C}{\longrightarrow} SU(3)_c \otimes
SU(2)_L \otimes U(1)_{R}
\otimes U(1)_{(B-L)} \stackrel{M_R}{\longrightarrow}
{\mathcal G}_{SM}.
\label{two-int}
\end{equation}
The Pati-Salam (${\mathcal
G}_{422} \equiv  SU(4)_c \otimes SU(2)_L
\otimes SU(2)_R$) decomposition gives:
\begin{equation}
45=(\Sigma_{3L},\Sigma_{3R},\Sigma_{4C},\Sigma_{LRC}) = 
(1,3,1)\oplus(1,1,3)\oplus(15,1,1)\oplus(6,2,2).
\end{equation}
It is useful to note the $SU(3)_c \otimes SU(2)_L \otimes U(1)_R \otimes 
U(1)_{B-L}$ decompositions
\begin{eqnarray}
(15,1,1)&\equiv&(1,1,0,0)+(3,1,0,-4/3)+(\bar{3},1,0,4/3)+(8,1,0,0)\;,\\
(4,1,2)&\equiv&(1,1,\pm\frac{1}{2},1)+(3,1,\pm\frac{1}{2},-1/3)\;.\nonumber
\end{eqnarray}

The colour, $U(1)_{R}$, and $U(1)_{(B-L)}$ singlet  members of $\Sigma_{3R}$
and $\Sigma_{4c}$ couple to $\nu_R$ when $\overline{16}_H$ gets a
{\em vev} along $(1,1,-\frac{1}{2},1)\subset$(4,1,2) that breaks
$U(1)_R \otimes U(1)_{B-L}$. The relevant Yukawa coupling is:
\begin{equation}
Y_{16}16_F 45_F \overline{16}_H \supset Y_{16} \left[ a_1 (1,1,\frac{1}{2},-1)_F
(1,1,0,0)_F^{\Sigma_{3R}}  + a_2 (1,1,\frac{1}{2},-1)_F (1,1,0,0)_F^{\Sigma_{4c}} \right]
(1,1,-\frac{1}{2},1)_H\;.
\end{equation}
$a_{1,2}$ are CG coefficients. The {\em vev} $v_R \equiv
<(1,1,-\frac{1}{2},1)_H>$ sets the scale $M_R$.

The masses of the adjoint matter fields are generated from
\begin{equation}
 M ~Tr(45_{F}^2) ~ + ~ \lambda ~Tr(45_{F}^2 210_{H})\;.
\label{mass45}
\end{equation}
Once $210_{H}$ acquires a {\em vev} along the (1,1,1) direction,
$SO(10)$ is broken to $SU(4)_c\otimes SU(2)_L\otimes SU(2)_R$. In
the mass term $M_{N}$ of $(1,1,0,0)_F$ $\subset$ $(15,1,1)_F$ and
$M_{\Sigma_{3R}}$ of $(1,1,0,0)_F$ $\subset$ $(1,1,3)_F$, an
extra contribution (from the second term of eq. (\ref{mass45}))
is added, {\em i.e.,}  $M_{\Sigma_{3R}}=M_{N}= M + \lambda
<(1,1,0,0)_H>$. There is no symmetry that protects the
masses of these adjoint fermions. So naturally these are very
heavy ($\sim M_X$).

\section{Constraints from  gauge coupling unification} 

In this section, we discuss the Renormalization Group (RG)
evolution of the gauge couplings at the one-loop level, check
for the scale of unification and determine the possible
intermediate scales. The symmetry breaks in two stages following
the steps given in (\ref{two-int}).  The contributions in the RG
running from scalars at the different scales  are included
according to the `extended survival hypothesis'\footnote{Only
those scalars are light which take part in the symmetry
breaking.} (ESH) \cite{ESH} which amounts to minimal fine tuning
of the parameters of the potential. Our model contains extra
adjoint fermions. But these fermions are very heavy $\sim
\cal{O}$ $(M_{X})$, so they do not contribute in the
renormalization group evolution of the gauge couplings.

When the $SO(10)$ symmetry is broken to the Pati-Salam group
\cite{Pati} ${\mathcal G}_{422}$ by a $210_H$ multiplet through
the {\em vev} in the  $<(1,1,1)>$ direction,
D-parity\footnote{D-parity is a symmetry that connects the
$SU(2)_L$ and $SU(2)_R$ sectors of a multiplet.}
\cite{D-parity} is spontaneously broken at this scale ($M_C$).  

\begin{table}[hbt]
\begin{center}
\begin{tabular}{|c|c|c|c|c|}
\hline
$SO(10)$&Symmetry&\multicolumn{3}{|c|}
{Scalars contributing to RG evolution}\\ \cline{3-5}
representation & breaking & $M_Z \rightarrow M_R$ & $M_R \rightarrow M_C$
& $M_C \rightarrow M_X$ \\
 & & Under ${\mathcal G}_{SM}$ & Under ${\mathcal
G}_{3211}$ & Under ${\mathcal
G}_{422}$   \\ \hline
{\bf 10} &  &  &  & (1,2,2)\\  
& ${\mathcal G}_{SM} \rightarrow EM$ & (1,2,$\pm1$)
&(1,2,$\pm\frac{1}{2}$,0) & \\
{\bf 120} &  &
...  & ... & (1,2,2), (15,2,2)\\  
& & & & \\
 {$\overline {\bf {16}}$} & ${\mathcal G}_{3211} \rightarrow {\mathcal G}_{SM}$ &
...  &  (1,1,$-\frac{1}{2}$,1) & (4,1,2) \\
& & & & \\
{\bf 210} & ${\mathcal G}_{422} \rightarrow {\mathcal G}_{3211}$ &
...& ...& (15,1,3) \\ \hline
\end {tabular}\\
\caption{Higgs submultiplets contributing 
to the RG evolution as per
the extended survival hypothesis when symmetry breaking of
$SO(10)$ takes place with two intermediate stages -- see (\ref{two-int}).
}
\label{t:esh10s2}
\end {center}
\end{table}

The gauge coupling evolution is usually stated as \cite{Georgi2}:
\begin{equation}
\mu \frac{dg_i}{d\mu} = \beta_i(g_i, g_j), \;\;\;\;\; (i,j = 1,\ldots,n),
\end{equation}
where $n$ is the number of couplings in the theory and at one-loop order
\begin{equation}
\beta_{i}(g_i, g_j)=(16\pi^{2})^{-1}b_{i}g_{i}^{3}.
\label{eq:RG}
\end{equation}
There is, however, a subtlety which must be taken into account
since the gauge symmetry in the energy range $M_R$ to $M_C$
includes two $U(1)$ factors. According to the ESH the $SO(10)$
multiplets are split in mass with some submultiplets having mass
above and some below this range. The incomplete scalar and
fermion multiplets that contribute to the RG evolution at this
stage lead to a mixing between these two $U(1)$ gauge groups.
Thus even at the one-loop level one cannot treat the evolution of
these $U(1)$ couplings in separation and in a generic scenario
one must include a (2 $\times$ 2) matrix of $U(1)$ couplings. The
details of this $U(1)$ mixing are skipped here\footnote {See, for
example, reference \cite{u1mixing}.}. We have computed
the RG-coefficients following the proposals given in
\cite{u1mixing} at the one-loop level including the $U(1)$
mixings. The $b_i$ are the ordinary beta-coefficients and the
$\tilde{b}_j$ are the additional ones which arise due to the
mixings stated above.  Taking all this into account, the gauge
couplings evolve as follows:
\noindent

{\bf i) From $M_C$ to $M_X$ }:
\begin{equation}
~~ b_{2L}=7/3; \;\;
b_{2R}=13; \;\;
b_{4c}=-1.
\label{xc}
\end{equation}

{\bf ii) From $M_R$ to $M_C$}:
\begin{equation}
~~ b_{2L}=-3; \;
b_{RR}=53/12; \;
b_{3c}=-7; \;
b_{(B-L)(B-L)}=33/8; \;
\tilde b_{R(B-L)}= \tilde b_{(B-L)R} = -1/4\sqrt{6}.
\label{mixing}
\end{equation}

{\bf iii) From $M_Z$ to $M_R$}:
\begin{equation}
~~ b_{1Y}=21/5; \;\;
b_{2L}=-3; \;\;
b_{3c}=-7.
\label{xc1}
\end{equation}
\begin{center}
\begin{figure}[thb]
\hskip 4.30cm
\includegraphics[width=8.5cm,height=6.50cm,angle=0]{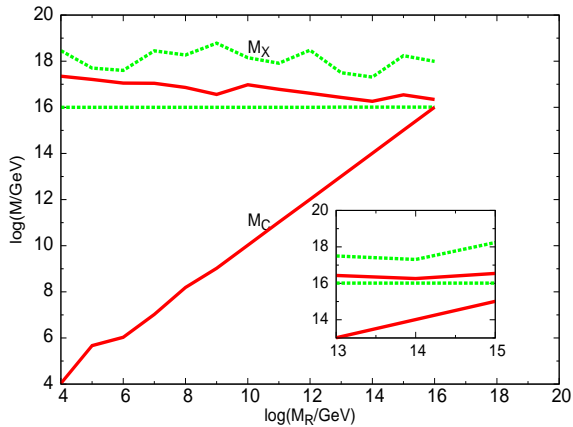}
\caption{\sf \small The allowed ranges of the unification ($M_X$,
pale, green) and intermediate Pati-Salam ($M_C$, dark, red) scales
as a function of the $U(1)_{(B-L)}$ breaking scale ($M_R$) for
$SO(10)$ with two intermediate scales.  The inset is a zoom of
the region of interest for generating  neutrino masses of the
right magnitude.}
\label{f:sv1sn}
\end{figure}
\end{center}
The mixing of the two $U(1)$ groups adds flexibility to the
model. With this, we find for every $M_R$ a range of consistent
solutions for $M_C$ and $M_X$ (see Fig. \ref{f:sv1sn}).   In the
plot we have exhibited the maximum and minimum values of both
$M_C$ and $M_X$ consistent with unification.  In a Grand Unified
Theory low intermediate scales are always perceived with extra
interest.  These low intermediate scale scenarios keep alive the
hope that signals of the GUT  may be identified at accessible
energies.  In Fig. \ref{f:sv1sn}, we have shown that $M_R$ and
$M_C$ can be quite low -- $\sim$ 10 TeV -- which is within the
reach of recent colliders, such as the LHC; this is  an artifact
of the inclusion of the $U(1)$ mixings.  The {\em vev}  $v_R$ of
the scalar  $(1,1,-\frac{1}{2},1)\subset\overline{16}$, sets the
scale $M_R$.  In the next section we have shown that $v_R$ needs
to be very high ($\sim 10^{14}$ GeV) to yield the correct
neutrino mass with the Yukawa couplings $\sim \cal O$(1). In the
inset of Fig. \ref{f:sv1sn} we magnify this range of $M_R$. It is
to be noted that this establishes that the proposed model of
`double-seesaw' mechanism is compatible with gauge coupling
unification at a scale which is not in conflict with the present
bound on the proton lifetime.

\section{Neutrino Mass} 
The neutrino mass matrix in the basis
($(\nu_{L})^{c},\nu_R,\Sigma^{0}_R,N$) is:
\begin{eqnarray}
\label{mnu}
M_\nu = \pmatrix{
  0 & m_D & 0 & 0
\cr
  m_{D}^{T}& 0 & a_1Y_{16}v_{R} &  a_2Y_{16}v_{R} 
\cr
0 & a_1Y_{16}^{T}v_{R} & M_{N} & 0 
\cr
 0 & a_2Y_{16}^{T}v_{R} & 0 & M_{N}
\cr}\;.
\end{eqnarray}
The left-handed fermionic triplets, $\Sigma_{3L}$, having a mass
matrix identical to $M_N$, do not mix  with other fermions since
the left-handed analogue of $v_R$ is chosen to be zero.  From the
mass matrix (\ref{mnu}) it is seen that the masses of the light
neutrinos are obtained by integrating out the heavy triplet and
singlet fermions.  Thus we can have type-III and type-I seesaw
mechanism in succession. The right-handed neutrino mass term is
generated once the heavy triplet fermion $\Sigma_{3R}^{0}$ and
$N$ are 
integrated out --  an effective type I + III seesaw.  Assuming $M_N
\gg ~
v_{R}Y_{16} \gg~ m_D$, the right-handed neutrino mass matrix is:
\begin{equation}
 M_{{R}}=v_{R}^2 Y_{16} M_M^{-1} Y_{16}^{T},
\label{e:MR}
\end{equation} 
where,
\begin{equation}
M_M^{-1} = (a_{1}^2 +a_{2}^{2}) M_{N}^{-1}\;\;,
\end{equation} 
and the light neutrino mass matrix after an effective type-I
seesaw becomes: 
\begin{equation}
m_{\nu}=m_{D} M_{R}^{-1} m_{D}^{T}\;\;.
\end{equation}
Substituting for $m_D$ from eq. (\ref{matrices}) one arrives
at the general expression of  $m_\nu$ as
\begin{equation} 
m_{\nu} = c_0^2 M_0  M_R^{-1} M_0 - c_0 c_4 M_0  M_R^{-1} M_2 
               + c_4 c_0 M_2  M_R^{-1} M_0 
               + c_4^2 M_2  M_R^{-1} M_2\;\;. 
\label{e:mnu1}
\end{equation} 
Typical values for the various parameters are 
$v_R \sim 10^{14}$ GeV, $ M_N \sim 10^{15}$ GeV, and
$c_{i} \sim {\cal{O}}(1)$, $Y_{i} \sim {\cal{O}}(1)$ which gives 
$M_R \sim 10^{12}$ GeV. Then with $m_D \sim 100$ GeV 
one gets $m_\nu \sim  1$ eV. 

With three neutrino generations, the model has 6 real parameters
in  $M_0$ and 3 in $M_2$.  In addition there are 5 {\em vev}s
($c_0,c_2,c_3,c_4,v_R$).  Besides, there are additional
parameters in  $Y_{16}$ and $M_N$.  However the low energy
neutrino mass matrix is characterized by 9 parameters. Neutrino
oscillation experiments have so far determined and/or bounded 5
of these. The general case is obviously not sufficiently constrained.
One way to address this lacuna requires invoking some flavour
symmetry.  We consider this to be the $\mu-\tau$ symmetry.

\section{$\mu-\tau$ symmetry and allowed textures} 

$\mu-\tau$ symmetry has been considered widely for explaining the
large atmospheric mixing angle in the neutrino sector
\cite{mu-tau}.  In addition it gives $\theta_{13} =0$ which is
also consistent with the current global fits\footnote{Recent
global fits have found indication for non-zero $\theta_{13}$
although this is only a 1$\sigma$ effect.  A small non-zero value
of $\theta_{13}$ can be induced  by breaking the $\mu-\tau$ symmetry.}.  We
impose the condition of a generalized $\mu-\tau$ symmetry on the
Yukawa matrices stemming from $10_H$ and $\overline{16}_H$.  This implies
that these matrices are invariant under the exchange of the
second and third rows and columns.  This reduces the number of unknown
parameters in the Yukawa sector.  However, this symmetry cannot
be exact in the quark and lepton sector.  This is accomplished by
the term $M_2$ in the fermion mass matrices which originates from
the $120_H$ which is taken to be antisymmetric under the exchange
of $2 \leftrightarrow 3$ and breaks $\mu-\tau$ symmetry
spontaneously.  
In
addition we had imposed a generalized parity symmetry
\cite{parity} which makes the complex matrices $M_0$ and $M_2$
real thereby reducing the number of free parameters.  Thus the
model that we consider is $SO(10) \otimes {Z_2}^{\mu-\tau}
\otimes Z_2^{\cal P}$ \cite{anjan-ketan}.  However it is to be
mentioned that if we assume exact $\mu-\tau$ (anti)symmetry  in
($M_2$) $M_0$ then a generalized CP-invariance holds
\cite{anjan-ketan} and the CKM matrix comes out as real.
This can be rectified either by assuming some of the {\em vev}s
to be complex or by allowing a small explicit breaking of
$\mu-\tau$ symmetry in $M_0$.  This induces CP-violation phases
in both $U_{CKM}$ and $U_{PMNS}$ \cite{anjan-ketan}.  We work in
the basis where the charged lepton mass matrix is diagonal and
the PMNS matrix is solely determined by the mixing in the
neutrino sector\footnote{For the purpose of this paper we only
consider the implications of this model for neutrino masses. The
predictions for the charged lepton and quark  masses would
require a detailed fit which we do not discuss in this work.}.

The structures for $M_0$ and $M_2$ under the above symmetries are
given by
\begin{eqnarray} 
M_0 = \left(
\begin{array}{ccc} 
a^\prime  & b^\prime & b^\prime \\ b^\prime & c^\prime & d^\prime
\\ b^\prime & d^\prime & c^\prime
\end{array} 
\right),
\quad
M_2 = \left(
\begin{array}{ccc} 
0  & x^\prime & -x^\prime \\ -x^\prime & 0 & y^\prime \\ x^\prime
& -y^\prime & 0
\end{array} 
\right) \;\;.
\label{M0M2}
\end{eqnarray}

We consider a model with three adjoint fermion multiplets, {\em i.e.,}
the model  consists of $(3 \nu_L + 3 \nu_R + 3 N + 3 \Sigma_R)$.
Thus, $Y_{16}$ and $M_N$ are also $(3 \times 3)$ matrices which we
take to be $\mu-\tau$ symmetric. It follows from eq. (\ref{e:MR})
that $M_R$ also respects this symmetry. 
Thus we have both $M_0$ and $M_R$ to be $\mu-\tau$ symmetric. 
In order to make the model predictive we make the further
assumption that $M_R$ and $M_0$ are proportional, {\em i.e.,}
\begin{equation} 
K M_R = M_0. 
\label{mrm0}
\end{equation} 
where $K$ is a constant. $m_\nu$ in eq. (\ref{e:mnu1}) then takes the form 
\begin{equation} 
m_{\nu} = K  c_0^2M_0  + K c_4^2M_2  M_0^{-1} M_2 
= M_{1} + M_{1}^\prime
  \;\;. 
\label{mnu1}
\end{equation}
The number of free real parameters in the theory are now 4 from
$M_0$, 2 in $M_2$, and 4 real {\em vevs}.  Because of
eq. (\ref{mrm0}) $M_R$ adds just one further parameter.  Thus in total
we have 11 real parameters.  The {\em vev} ratios $c_2$ and $c_3$
do not affect eq. (\ref{mnu1}) and thus we have 9 parameters
involved in the neutrino sector.  Some of these appear only
as overall scale factors.  

We note that although $M_2$  is $\mu-\tau$ antisymmetric the
product $M_2  M_0^{-1} M_2$ possesses $\mu-\tau$ symmetry.  Thus
$m_\nu$ is $\mu-\tau$ symmetric. This immediately implies
$\theta_{13} =0$ and $\theta_{23} = \pi/4$.  Therefore the mixing
matrix in the basis where the charged lepton mass matrix is
diagonal is given as,
\be
\label{upmns} 
U_{\rm PMNS} = \left(
\bad
c_{12} & s_{12} & 0  \\
-s_{12}/\sqrt{2} & c_{12}/\sqrt{2} & 1/\sqrt{2} \\
-s_{12}/\sqrt{2} & c_{12}/\sqrt{2} & -1/\sqrt{2} \\
\ea
\right).
\ee
which can be brought to the standard $U_{\rm PMNS}$ form by
a suitable redefinition of fermion phases.
We have 
\be
m_\nu = U_{\rm PMNS} M_{dia} U_{\rm PMNS}^T,
\label{mnusolve} 
\ee
where $M_{dia} = Diag(m_1,m_2,m_3)$.  $m_1,m_2,m_3$, the mass
eigenvalues are real \footnote{Since the mass matrices have real entries,
complex roots can appear only in conjugate pairs leading to unacceptable
degenerate neutrinos. We take the eigenvalues to be all non-negative.}, 
and are given as
\be
m_1  = \frac{ X  -  \sqrt{X^2 - 4 (d - c ) Y}}{2(d-c )}, ~~ 
m_2  =  \frac{ X  +  \sqrt{X^2 - 4 (d - c ) Y }}{2(d-c )},~~ 
m_3  =   \frac{Y}{2 b^2 - a c - a d}.
\ee
Here
\begin{eqnarray}
X & = & -a c - c^2 + a d + d^2 + 2 x^2 + y^2; 
\nonumber \\ 
Y & = & 2 b^2 c - a c^2 - 2 b^2 d + a d^2 + 2 c x^2 + 2 d x^2 + 4
b x y + a y^2, 
\end{eqnarray}
and
\be
a = K c_0^2 a^\prime,~~ b = K c_0^2 b^\prime,~~ c = K c_0^2
c^\prime,~~ d = K c_0^2 d^\prime,~~ x = K c_4^2 x^\prime,~~ y = K
c_4^2 y^\prime \;\;.
\label{abcd} 
\ee

Note that the eigenstate $m_3$ is determined to be the 
one associated with the eigenvector $(0,1/\sqrt{2}, -1/\sqrt{2})$.
Whether this is the highest mass state or the lowest mass state {\it {i.e.}}
whether 
the  hierarchy  is normal or inverted will depend on 
the values of the parameters. 
We further require $\Delta m^2_{21} >0$ from the solar data. 
This implies that for our choice of $m_2$ and $m_1$ 
\be
 \frac{ X}{(d-c )^2}  
     \sqrt{X^2 -
      4 (d - c ) Y }~~>~~ 0
\ee
Using eqs. (\ref{upmns}) and (\ref{mnusolve}) 
we obtain, 
\be
\tan\theta_{12} = \frac{1}{\sqrt{2}} \frac{(a- m_1)(c-d) - 2x^2
}{b (c-d) + xy}\;\;.
\ee
The condition for tri-bimaximal mixing implies
\be
(a-m_1 - b)(c-d)  =  2 x^2 + xy \;\;.
\ee

\subsection{$10_H$ dominance} 

In this case, $a, b, c, d \gg x, y$. The light neutrino mass
matrix $m_\nu$ is approximated as $K c_0^2 M_0$  with $M_0$ defined in
eq. (\ref{M0M2}). In this limit the mass eigenvalues are given
as,
\be
m_1  = \frac{1}{2} (f_1 - R) ,\,\, 
m_2 = \frac{1}{2}(f_1 + R) ,\,\,
m_3 = c-d  \;\;,
\label{eq:mnueigen2}
\ee
with 
\be
R = + \sqrt{8 b^2 + f_2^2}  \;\;,
\ee 
where, 
\be
f_1 = a+c+d, ~~f_2 = -a+c+d  \;\;.
\label{abcdnew} 
\ee

Again, $m_3$ is identified as the eigenvalue for the state  
eigenvector $(0,1/\sqrt{2},-1/\sqrt{2})$.  
Since the solar data has determined the ordering of the 1 and 2 mass states to 
Then the mass squared differences can be expressed as, 
\be
\Delta m^2_{21} = f_1 R 
\qquad \Delta m^2_{31} = (f_1 R - a^2 - 4 b^2 + c^2 + d^2 - 6cd)/2  \;\;.
\ee
Again, the mass ordering will  depend on the values of the parameters. 
In general both normal and inverted hierarchy 
are possible.  In addition, the solar neutrino data require
$\Delta m^2_{21} > 0$  which implies $f_1 R >0$ for the above 
selection of states. 

The mixing angles are given as, 
\be
\theta_{13}^\nu = 0
~,\;\;
\theta_{23}^\nu =  \pi/4
~,\;\;
\tan\theta_{12}^\nu 
=\frac{(R-f_2)}{2\sqrt{2}b}
~.
\label{eq:mixnu}
\ee

Tri-bimaximal mixing implies $\theta_{13} =0$, $\theta_{23} =
\pi/4$ and $\tan^2\theta_{12} = 1/2$.  We see that the
requirements for $\theta_{13}$  and $\theta_{23}$ are already satisfied.
If in addition we impose
\be
 f_2 = b ~~\Longrightarrow ~~R = 3b, ~f_1 = (2a+b),
\ee
tri-bimaximal mixing is obtained. In this limit   
\be
\Delta m^2_{21} = 3~b~(2a + b) 
\qquad \Delta m^2_{31} = (c - d)^2 - (a - b)^2 \;\;.
\ee



\subsection{$120_H$ dominance} 

In this limit $a, b, c, d \ll x, y$ and the low energy neutrino
mass matrix is given as
\be
m_\nu = M_4 = K c_4^2 M_2~ M_0^{-1}~ M_2  \;\;.
\ee

The $U_{\rm PMNS}$ continues to be given by eq. (\ref{upmns}).  
The eigenvalues, in terms of the parameters defined in eq. (\ref{abcd}), 
are given as, 
\be
m_1 = 0 ,\,\,
m_2 =  \frac{2 x^2 +y^2}{d-c} ,\,\, 
m_{3} =  \frac{2 c x^2 + 2 d x^2 + 4 b x y + a y^2}
{2 b^2 - a c - a d
} \;\;.
\ee
Since the eigenvector $(0,1/\sqrt{2},-1/\sqrt{2})$  
belongs to the eigenvalue $m_3$ so that the zero eigenvalue has to be 
associated with the eigenstate $m_1$.  
Therefore this case corresponds to 
the normal hierarchy.  
Since $m_1 =0$, $\Delta m^2_{21} = m_2^2$ and 
$\Delta m^2_{31} = m_3^2$.  
Then, using eqs. (\ref{upmns}) and (\ref{mnusolve}) one obtains
the 1-2 mixing angle as,
\begin{equation}
{\tan\theta_{12}}  =   -\frac{\sqrt{2} x}{y}
\end{equation} 
Thus, the mixing matrix in this case is completely determined by 
the parameters of $M_2$.  
The condition for obtaining exact tri-bimaximal mixing is 
$y = -2x$.



\section{Conclusions}

We consider a non-SUSY $SO(10)$ model in which the fermion masses originate 
from Yukawa couplings to $10_H$ and $120_H$. In such a model the 
usual type-I and type-II seesaw mass terms which originate from
$\overline{126}_H$ are not present. 
Here, it is possible to generate the  neutrino mass at two loops 
by the radiative seesaw mechanism
\cite{witten}. But for non-SUSY $SO(10)$ the contribution is very small. 

In this paper we suggest a new possibility to generate neutrino
masses in a non-SUSY $SO(10)$ model with $10_H + 120_H$ using
fermions in the $45_F$ representation and an additional
$\overline{16}_H$ scalar multiplet. Constraints from gauge
coupling unification requires these $vev$ $<\overline{16}_H>$ to be in the
range $ \sim 10^{4} - 10^{16}$ GeV. However from the standpoint
of generation of naturally small neutrinos masses the range $\sim
10^{13} - 10^{15}$ GeV is preferred.  We show that in this case
one can generate small neutrino masses through the `double
seesaw' mechanism. Predictions for mixing angles require further
imposition of a flavour symmetry which we chose to be the
$\mu-\tau$ symmetry for the Yukawa matrices due to $10_H$ and
$\overline{16}_H$ whereas for the one originating from $120_H$ we assume the
matrix to be $\mu-\tau$ antisymmetric. We further assume the
right-handed matrix ($M_R$) due to the heavy fields to be proportional to
the one ($M_0$) originating from $10_H$.  With this the light
neutrino mass matrix is given by the sum of two terms  which are
both $\mu-\tau$ symmetric. This automatically satisfies
$\theta_{13} = 0$ and $\theta_{23} = \pi/4$.  We present the
neutrino masses and $\theta_{12}$ obtained from this model and
determine the condition for satisfying tri-bimaximality. We also
discuss the limiting values when one of the terms
dominate. For the $10_H$-dominance case  both hierarchies are possible whereas
if the $120_H$ dominates the hierarchy can only be normal.

\section{Acknowledgement} 
S.G. wishes to thank Anjan Joshipura for many useful discussions
and acknowledges the hospitality at Harish-Chandra Research Institute
during the course of this work. 
This research has been supported by funds from the XIth Plan
`Neutrino Physics' and RECAPP projects at Harsih-Chandra Research Institute.
AR acknowledges
partial support from a J.C. Bose Fellowship of the Department
of Science and Technology.


\begin{thebibliography}{99}

\bibitem{thomas} 
  T.~Schwetz, M.~A.~Tortola and J.~W.~F.~Valle,
  New J.\ Phys.\  {\bf 10} (2008) 113011
  [arXiv:0808.2016 [hep-ph]].




\bibitem{tribi}
  P.~F.~Harrison, D.~H.~Perkins and W.~G.~Scott,
  Phys.\ Lett.\  B {\bf 530} (2002) 167
  [arXiv:hep-ph/0202074].

\bibitem{Georgi1}
  H.~Georgi and S.~L.~Glashow,
  Phys.\ Rev.\ Lett.\  {\bf 32} (1974) 438.
 


\bibitem{Georgi2}
  H.~Georgi, H.~R.~Quinn and S.~Weinberg,
  Phys.\ Rev.\ Lett.\  {\bf 33} (1974) 451.

\bibitem{Pati} J. C. Pati and A. Salam, Phys. Rev. Lett. {\bf 31} (1973) 66;
 Phys. Rev. {\bf D 10} (1974) 275.

\bibitem{so10}
H. Georgi, {\em In Coral Gables 1979 Proceeding, Theory and experiments
in high energy physics,} New York 1975, 329;
H.~Fritzsch and P.~Minkowski,
Annals Phys.\  {\bf 93} (1975) 193.

\bibitem{Seesaw}
P.~Minkowski,
Phys.\ Lett.\ B {\bf 67} (1977) 421; 
T.~Yanagida, Proceedings of the {\em Workshop on Unified Theories
and Baryon Number in the Universe}, Tsukuba, 1979, eds.  A.
Sawada, A. Sugamoto; 
S.~Glashow, in {\em Carg\`ese 1979, Proceedings, Quarks and
Leptons} NATO Adv.  Study Inst. Ser.B Phys. {\bf 59} (1979) 687;
M.~Gell-Mann, P.~Ramond and R.~Slansky, Proceedings of the {\em
Supergravity Stony Brook Workshop}, New York, 1979, eds. P. Van
Niewenhuizen, D. Freedman {\it et al.} (North-Holland, Amsterdam,
1980);
R.~Mohapatra and G.~Senjanovi\'c, Phys. Rev. Lett. {\bf 44} (1980)
912.  

\bibitem{goran} 
  G.~Senjanovi\'{c},
  arXiv:hep-ph/0612312.

\bibitem{babu-mohapatra} 
  J.~A.~Harvey, D.~B.~Reiss and P.~Ramond,
  Nucl.\ Phys.\  B {\bf 199} (1982) 223;
  G.~Lazarides and Q.~Shafi,
  Nucl.\ Phys.\  B {\bf 350} (1991) 179;
  K.~S.~Babu and R.~N.~Mohapatra,
  Phys.\ Rev.\ Lett.\  {\bf 70} (1993) 2845
  [arXiv:hep-ph/9209215];
  C.~H.~Albright and S.~Nandi,
  Phys.\ Rev.\ Lett.\  {\bf 73} (1994) 930
  [arXiv:hep-ph/9311227].


\bibitem{seesaw2} 
M.~Magg and C.~Wetterich,
Phys.\ Lett.\  B {\bf 94}, (1980) 61;
G.~Lazarides, Q.~Shafi and C.~Wetterich,
Nucl.\ Phys.\  B {\bf 181}, (1981) 287.

\bibitem{10+126-susy} 
  K.~y.~Oda, E.~Takasugi, M.~Tanaka and M.~Yoshimura,
  Phys.\ Rev.\  D {\bf 59} (1999) 055001
  [arXiv:hep-ph/9808241]; 
  H.~S.~Goh, R.~N.~Mohapatra and S.~P.~Ng,
  Phys.\ Rev.\  D {\bf 68} (2003) 115008
  [arXiv:hep-ph/0308197];
  H.~S.~Goh, R.~N.~Mohapatra, S.~Nasri and S.~P.~Ng,
  Phys.\ Lett.\  B {\bf 587} (2004) 105
  [arXiv:hep-ph/0311330];
  S.~Bertolini, M.~Frigerio and M.~Malinsky,
  Phys.\ Rev.\  D {\bf 70} (2004) 095002
  [arXiv:hep-ph/0406117];
  T.~Fukuyama, A.~Ilakovac, T.~Kikuchi, S.~Meljanac and N.~Okada,
  Eur.\ Phys.\ J.\  C {\bf 42} (2005) 191
  [arXiv:hep-ph/0401213]; 
  B.~Bajc, A.~Melfo, G.~Senjanovi\'{c} and F.~Vissani,
  Phys.\ Rev.\  D {\bf 70} (2004) 035007
  [arXiv:hep-ph/0402122];
  C.~S.~Aulakh and A.~Girdhar,
  Nucl.\ Phys.\  B {\bf 711} (2005) 275
  [arXiv:hep-ph/0405074].

\bibitem{10+126-nosusy} 
  B.~Bajc, A.~Melfo, G.~Senjanovi\'{c} and F.~Vissani,
  Phys.\ Rev.\  D {\bf 73} (2006) 055001
  [arXiv:hep-ph/0510139].

\bibitem{allthree} 
  W.~Grimus and H.~K\"uhb\"ock,
  Phys.\ Lett.\  B {\bf 643} (2006) 182
  [arXiv:hep-ph/0607197];
  B.~Dutta, Y.~Mimura and R.~N.~Mohapatra,
  Phys.\ Lett.\  B {\bf 603} (2004) 35
  [arXiv:hep-ph/0406262].

\bibitem{anjan-ketan}
  A.~S.~Joshipura, B.~P.~Kodrani and K.~M.~Patel,
  Phys.\ Rev.\  D {\bf 79} (2009) 115017
  [arXiv:0903.2161 [hep-ph]].

\bibitem{10+120} 
  K.~Matsuda, Y.~Koide and T.~Fukuyama,
  Phys.\ Rev.\  D {\bf 64} (2001) 053015
  [arXiv:hep-ph/0010026].

\bibitem{Lavoura:2006dv}
  L.~Lavoura, H.~K\"uhb\"ock and W.~Grimus,
  Nucl.\ Phys.\  B {\bf 754} (2006) 1
  [arXiv:hep-ph/0603259].

\bibitem{witten} 
  E.~Witten,
  Phys.\ Lett.\  B {\bf 91} (1980) 81.



\bibitem{goran-bajc-prl} 
  B.~Bajc and G.~Senjanovi\'{c},
  Phys.\ Rev.\ Lett.\  {\bf 95} (2005) 261804
  [arXiv:hep-ph/0507169].

\bibitem{goran-split} 
  B.~Bajc and G.~Senjanovi\'{c},
  Phys.\ Lett.\  B {\bf 610} (2005) 80
  [arXiv:hep-ph/0411193].




\bibitem{double-seesaw} 
R.~N.~Mohapatra,
  Phys.\ Rev.\ Lett.\  {\bf 56} (1986) 561; 
S.~M.~Barr,
  Phys.\ Rev.\ Lett.\  {\bf 92} (2004) 101601
  [arXiv:hep-ph/0309152]; 
  S.~M.~Barr and I.~Dorsner,
  Phys.\ Lett.\  B {\bf 632} (2006) 527
  [arXiv:hep-ph/0507067].
 
\bibitem{TypeIII}
  R.~Foot, H.~Lew, X.~G.~He and G.~C.~Joshi,
  Z.\ Phys.\ C {\bf 44} (1989) 441.

\bibitem{perez-su5}
  P.~Fileviez P\'{e}rez,
  Phys.\ Rev.\  D {\bf 76} (2007) 071701
  [arXiv:0705.3589 [hep-ph]];
  I.~Dorsner and P.~Fileviez P\'{e}rez,
  JHEP {\bf 0706} (2007) 029.

\bibitem{goran-triplet}
  B.~Bajc, M.~Nemevsek and G.~Senjanovi\'{c},
  Phys.\ Rev.\  D {\bf 76} (2007) 055011
  [arXiv:hep-ph/0703080];
  A.~Arhrib, B.~Bajc, D.~K.~Ghosh, T.~Han, G.~Y.~Huang, I.~Puljak and G.~Senjanovi\'{c},
  Phys.\ Rev.\  D {\bf 82} (2010) 053004
  [arXiv:0904.2390 [hep-ph]].


\bibitem{perezlr}
  P.~Fileviez P\'{e}rez,
  JHEP {\bf 0903} (2009) 142
  [arXiv:0809.1202 [hep-ph]].


\bibitem{mu-tau}
  C.~S.~Lam,
  Phys.\ Lett.\  B {\bf 507} (2001) 214
  [arXiv:hep-ph/0104116].

\bibitem{mohapatra-nasri}
  R.~N.~Mohapatra, S.~Nasri and H.~B.~Yu,
  Phys.\ Lett.\  B {\bf 636} (2006) 114
  [arXiv:hep-ph/0603020].


\bibitem{parity} 
  W.~Grimus and H.~K\"uhb\"ock,
  Eur.\ Phys.\  J.\ C {\bf 51} (2007) 721 
  [arXiv:hep-ph/0612132].



  


\bibitem{ESH}
  F.~del Aguila and L.~E.~Iba\~nez,
  Nucl.\ Phys.\  B {\bf 177} (1981) 60.


\bibitem{D-parity}
  D.~Chang, R.~N.~Mohapatra and M.~K.~Parida,
  Phys.\ Rev.\ Lett.\  {\bf 52} (1984) 1072.


\bibitem{u1mixing}
B.~Holdom,
  Phys.\ Lett.\  B {\bf 166} (1986) 196; 
F.~del Aguila, G.~D.~Coughlan and M.~Quiros,
  Nucl.\ Phys.\  B {\bf 307} (1988) 633
  [Erratum-ibid.\  B {\bf 312} (1989) 751]; 
 S.~Bertolini, L.~Di Luzio and M.~Malinsky,
  Phys.\ Rev.\  D {\bf 80} (2009) 015013
  [arXiv:0903.4049 [hep-ph]]; 
J.~Chakrabortty and A.~Raychaudhuri,
  Phys.\ Rev.\  D {\bf 81} (2010) 055004
  [arXiv:0909.3905 [hep-ph]].


\end{thebibliography}
\end{document}